\documentclass[aps,prl,twocolumn]{revtex4}
\usepackage{graphicx}
\usepackage{color}
\begin{document}
\title{On the birefringence of healthy and malaria-infected red blood cells}
\author{Aditya K. Dharmadhikari, Himanish Basu, Jayashree A. Dharmadhikari, Shobhona Sharma, and Deepak Mathur{\footnote{email: atmol1@tifr.res.in}}}

\affiliation{Tata Institute of Fundamental Research, 1 Homi Bhabha Road, Mumbai 400 005, India}


\begin{abstract}
We have probed how the birefringence of a healthy red blood cell (RBC) changes as it becomes infected by a malarial parasite. By analyzing the polarization properties of light transmitted through a single, optically-trapped cell we demarcate two types of birefringence: form birefringence which depends on the shape of the cell and intrinsic birefringence which is brought about by the presence of the parasite. We quantitatively measure changes in the refractive index as normal RBS become infected by a malarial parasite. Malarial infections are found to induce changes in the cell's refractive index whose magnitude depends on the stage of malarial infection; such changes were quantitatively explored and found to be large, in the range 1.2 to 3$\times$10$^{-2}$. Our results have implications for the development and use of non-invasive techniques that seek to quantify changes in cell properties induced by pathological states accompanying diseases like malaria. From a broader prespective, information forthcoming from our trap-based experiments may stimulate physicists to initiate theoretical and computational efforts that have a bearing on cell biology issues related to diagnosis of diseases and therapeutics.
\end{abstract}
\maketitle

The importance of blood has been known to humankind since ancient Indian texts on internal medicine were written around 300 BCE \cite{Valiathan} and, subsequently, the human circulatory system was described by the Egyptian surgeon, Ibn An-Nafis \cite{Haddad, Meyerhof}, in the 13th century.  Optical physics intervened in the 17th century with microscopes enabling direct visualization of single red blood cells (RBCs) from 1658 onwards \cite{Hajdu}. Contemporary development of optical tools, like phase contrast and topographic phase microscopies \cite{choi, philips}, has opened newer vistas for studies of the biomechanical properties of single cells. Blood cell properties, including mechanical properties like membrane elasticity that affect blood flow through capillaries, are known to be affected by diseases like malaria. In spite of {\em Plasmodium falciparum} malaria being a prevalent pathological parasitic disease world-wide the exact events leading to severe pathology associated with it remain unclear. A decrease in RBC deformability is suspected to be the main culprit, with changes in cell rigidity being predicted to lead to improper flow of RBCs. This allows parasitized RBCs to block micro-capillaries, curtailing blood supply to the brain and other vital organs. Thus, quantitative information on the biomechanical properties of malaria parasite infected cells is likely to be important from the viewpoint of gaining insights into disease pathology. The adoption of dipole trap technology enables such information to be obtained on a single-cell basis. Here we describe utilization of a dipole trap \cite{Ashkin, currsci} to quantitatively monitor the influence that a malarial parasite has on optical properties of RBCs; specifically, we probe the birefringence that is induced in the cell by the malarial parasite. It has been known for more than half a century that healthy human RBCs exhibit intrinsic birefringence \cite{Mitchison, Ponder, hofrichter} which arises from the cell's hemoglobin \cite{Ponder, hofrichter} and membrane \cite{Mitchison}. Upon infection by a malarial parasite, the cell's hemoglobin content undergoes changes that affect birefringence: as the parasite grows (on time scales of hours), it crystallizes hemoglobin into hemozoin \cite{coban, romagosa}. Hemozoin crystals are highly birefringent and they, thus, offer opportunities for studying malaria-infected RBCs through polarization microscopy \cite{romagosa}. However, infection of an RBC by Plasmodium falciparum also results in significant shape alterations, with the shape anisotropy contributing to the cell's shape birefringence. By quantifying birefringence in normal (healthy) but shape-distorted cells, and in parasitized cells at different stages of infection, we succeed in distinguishing between form and intrinsic birefringence. The former depends on cell shape whilst the latter is induced by the malarial parasite. We also quantitatively determine changes in the refractive index of normal and infected RBCs by analyzing the polarization properties of light transmitted through single, optically-trapped cells; the refractive index can more than double at different stages of malarial infection. Our systematic experiments on how matter affects light were conducted on live cells maintained under physiological conditions; our findings have implications for the development and use of non-invasive methods for quantifying changes in cell properties induced by pathological states that accompany diseases like malaria. From a broader perspective, information forthcoming from our trap-based experiments may also stimulate physicists to initiate theoretical and computational efforts that have a bearing on cell biology issues related to diagnosis of diseases, therapeutics and general human health problems. 

It has been established in earlier experiments \cite{OE1} that malaria-infected RBCs undergo shape distortion and rotational motion upon their interaction with a tightly focused laser beam of the type that is encountered within a dipole trap. The distortion of shape has been characterized in terms of Euler buckling \cite{Physbio}; it plays an important role in the rotational motion by enhancing the spatial anisotropy of the cell's polarizability tensor, $\alpha$, such that the  {($\alpha$.E)} interaction with the optical field gives rise to a torque that, in turn, causes rotation of the trapped RBC ($E$ denotes the amplitude of the laser's optical field). The torque is a consequence of the unbalanced optical forces acting on different sides of the trapped cell. Rotations due to such shape anisotropy are expected to be similar to those of RBCs that have undergone hyper-osmotic shock \cite{JBOkapil} (the osmotic shock also induces shape changes); they scale with laser power. Such rotations have been experimentally characterized in recent years by combining optical trap technology with fluorescence and liquid flow methods \cite{JBOkapil} but a proper insight into the physics that drives the dynamics remains elusive. Moreover, rotations are also known to be caused in optically trapped objects that may be isotropic in shape but are {\em optically anisotropic}, as in birefringent materials that possess two axes of polarization, with one axis being more polarized than the other. Upon placing such a birefringent material in an optical trap, rotation is induced such that the laser light passes through the more polarizable axis \cite{block, dunlop}.

\begin{figure}
\includegraphics[width=10cm]{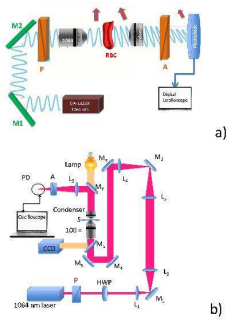}
\caption{(Color online) a) Schematic representation of how the plane of polarization of light changes after passing through a birefringent material (like an RBC). The transmitted light is analyzed by analyzer A, and is collected on a photodiode. b) Our optical trap set-up including a half-wave plate polarizer (HWP) and analyzer (A). The leaked light is collected on a photodiode connected to an oscilloscope.}
\end{figure}

In order to quantitatively probe birefringence in RBCs, we trapped and rotated cells located between a crossed polarizer-analyzer combination and recorded the intensity of the emerging light (see Fig. 1). With no RBC present, the light emerging out of the analyzer was negligibly small, as its plane was crossed with the plane of the polarizer. However, the presence of an RBC ensures a change in the angle of polarization of the light after it passes through the polarizer, with this change being dependent of the angle of polarization of the incident light and the angle of the RBC (Fig. 1a). As the cell rotates, this angle continuously changes, with the maximum change in the angle of polarization occurring when the angle between the RBC's axis and the angle of the incident light is 45$^o$ (Fig. 1b). The value at the maxima that is observed under such circumstances is a measure of the RBC's intrinsic birefringence. 

\begin{figure}
\includegraphics[width=10cm]{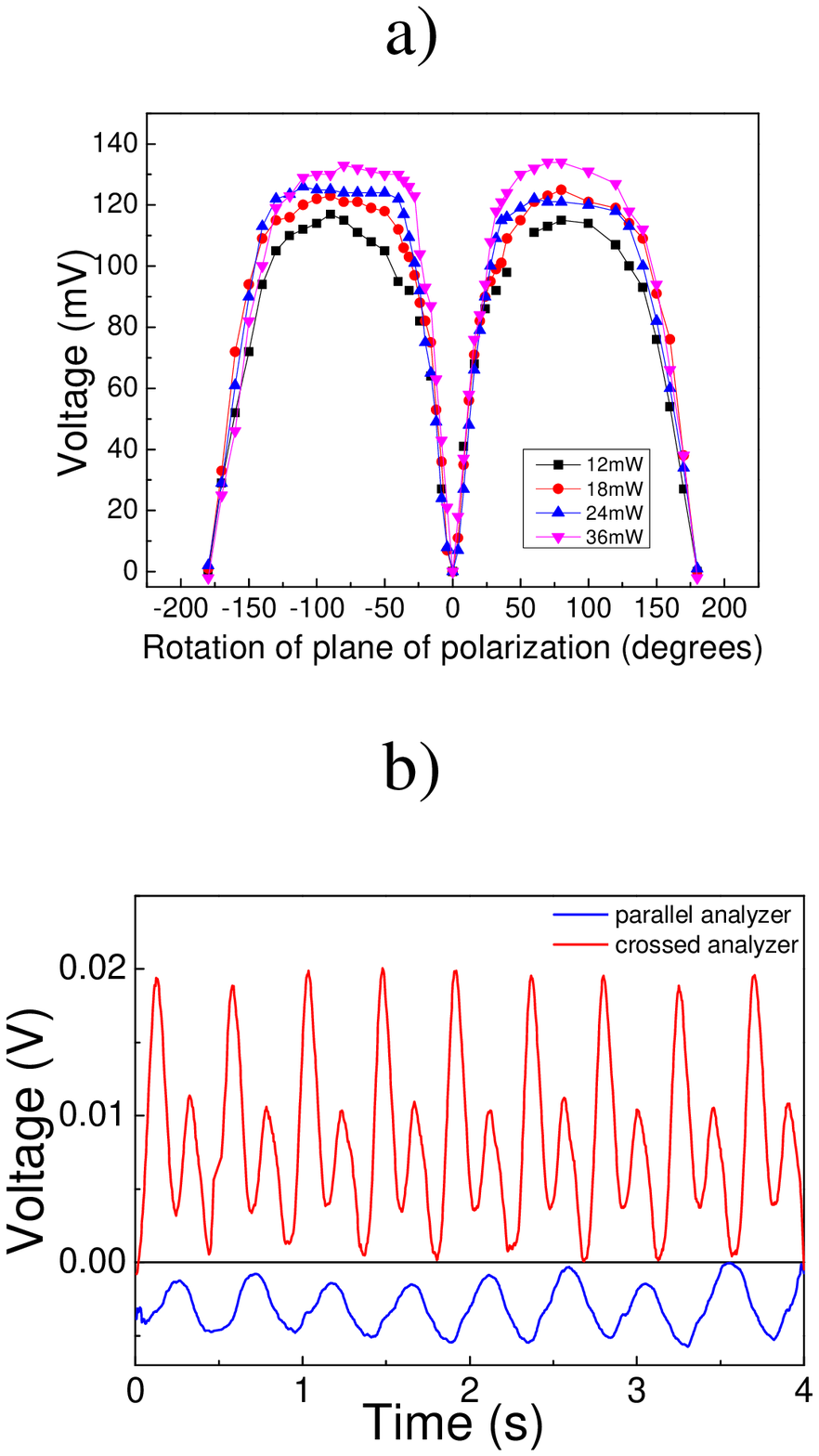}
\caption{(Color online) a) Voltage obtained on the photodiode due to light leakage in between a crossed-polarizer and analyzer, when the plane of polarization of the light after the polarizer is changed by means of a half-wave plate. b) Rotation spectra of RBC recorded on a photodiode when the analyzer is crossed with the polarizer (red line) and when the analyzer has its plane parallel to the polarizer (blue).}
\end{figure}

Figure 1a is a schematic illustration of how a birefringent material changes the plane of polarization of incident light that is plane-polarized. To probe the RBC birefringence we utilized the set-up shown in Fig 1b. Here, we modified an optical trap to incorporate a polarizer (to make the trapping laser light plane-polarized), a half-wave plate (to change the plane of polarization), and an analyzer to probe birefringence of trapped cells. Light from the half-wave plate passed through the microscope objective to focus on a single RBC. Since RBC are known to be birefringent, we expect the transmitted light to have its plane of polarization rotated by some angle and, consequently, the analyzer is expected to ``leak" some light. This light is then collected by a photodiode and is monitored on an oscilloscope. In order to first calibrate our set-up we crossed the polarizer-analyzer such that no light was transmitted: this defined zero degree. As the half-wave plate (HWP) was rotated, some leaked light started impinging on the photodiode. In Fig. 2a we plot typical photodiode voltages measured as the HWP was rotated. The effect of laser power on the rotation angle was found to be negligible. We thus established a one-to-one relation between the photodiode voltage and the degree of rotation. Hence, any rotation induced by birefringence could now be quantified using our optical trap: for any rotational motion of an RBC trapped in cross-polarized light the photodiode voltage is a direct measure of the angle by which the plane of polarization is rotated as light passes through the birefringent RBC. This angle is related to the change in refractive index. Figure 2b shows a typical plot. Occurrence of two peaks when the analyzer is crossed in a single rotation cycle is clear-cut indication of the trapped RBC's birefringence. When the analyzer is parallel we observe only a single peak, indicating a chopper-like behavior of the RBC.

\begin{figure}
\includegraphics[width=10cm]{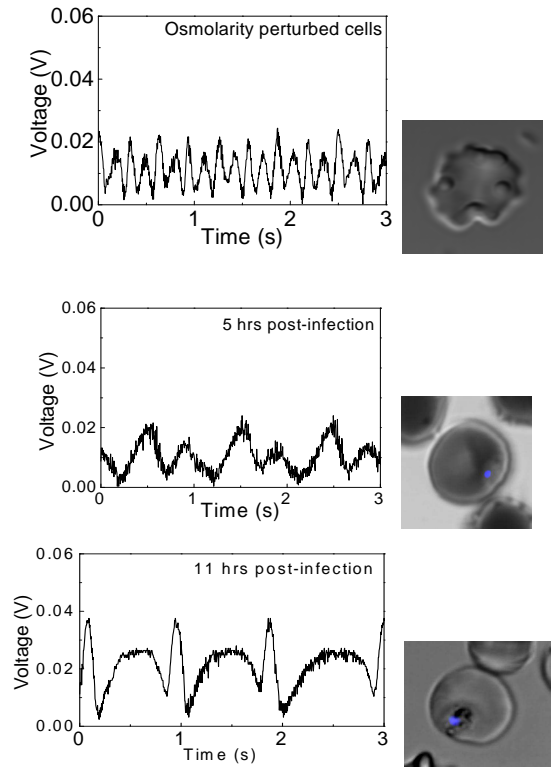}
\caption{(Color online) Rotation of RBCs as recorded on a photodiode. The right panel shows a DIC image of the RBC stained with a dye (DAPI) that highlights the malarial parasite.}
\end{figure}

We have carried out measurements on two types of shape-distorted RBCs: those whose shapes are distorted simply by changing their osmolarity, and those whose shapes become distorted because of the malarial parasite within them. As already noted, there are two types of anisotropies: those which is related to shape changes - they give rise to form birefringence - and those which are induced by biochemical alterations within the cell - these give rise to intrinsic birefringence. In case of osmolarity-induced shape-distorted RBCs, the resulting form birefringence induces the cell to rotate within our trap. Fig. 3 shows results of measurements carried out on such cells. Also shown are results of rotations caused by the intrinsic birefringence of malaria-infected cells. In the former case we made measurements on cells using osmolarities ranging from 150 oSm to 1200 oSm; the rotational behavior did not change with change in osmolarity, with the peak-to-peak photodiode voltage remaining at 20 mV. The angle of rotation measured for 20 mV change in photodiode voltage corresponds ~6$^0$ which when put in the equation \cite{hecht}
\[2\Delta\theta=\Delta n L 2 \pi / \lambda\] (where $\Delta$n is the refractive index change) yields a value of $\Delta$n to be 1.2$\times$10$^{-2}$ for $L=4\mu m$ and wavelength $\lambda$=1064 nm. In contrast, malaria-infected cells yielded a photodiode voltage that depended on the stage of infection. We made measurements with a synchronous culture of the malarial parasite at fixed time points. Data obtained after 5 hours and 11 hours of infection is presented in Fig. 3 and clearly shows that the photodiode voltage increases as the malarial stage progresses, indicating stage-dependent change in the infected RBC's intrinsic birefringence. In the early stages of infection (5h) the peak-to-peak voltage is 20 mV; this doubles 11h post infection, with the corresponding $\Delta$n change of 1.9$\times$10$^{-2}$. Twenty hours post infection $\Delta$n becomes 3$\times$10$^{-2}$. These values are consistent with the one previous microscopy report on malaria-infected RBCs \cite{suresh}. 

It is noteworthy that there is a difference in the rotation observed in osmolarity-perturbed cells and those that are malaria-infected. In former, the periodic rotations are symmetric whereas for malaria-infected cells they are distinctly asymmetric. Also, within a single cycle of rotation, there is marked asymmetry in the voltage value; this indicates that there is an inherent ellipticity in both infected and osmolarity-altered cells. 

\begin{figure}
\includegraphics[width=10cm]{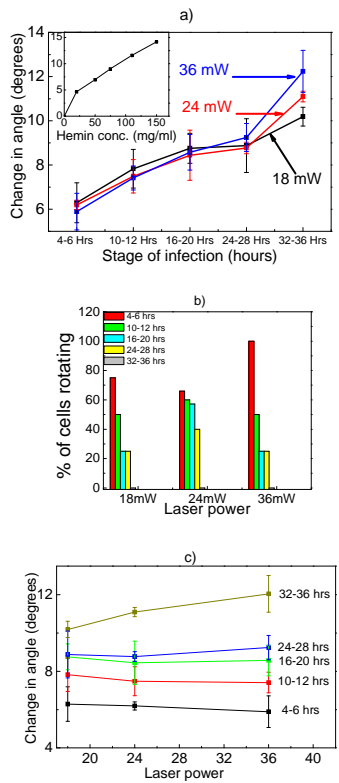}
\caption{(Color online) a) Change in the angle of polarization of the light that is induced by infected RBCs at various laser power levels. b) Percentage of cells that rotate at various stages of infection. c) change in plane of polarization as a function of laser power.}
\end{figure}

In Fig. 4a we show the change in angle as a function of infection stage for different laser powers. As the infection progresses the angle also increases, highlighting the additional contribution to birefringence being made by parasite within the cell. This additional contribution is likely to be due to increased hemin concentration. To verify this conjecture we carried out measurements using neat hemin solution (at varying concentrations); we found that for hemin concentration of 120 mg/ml the change in angle (10$^o$) is comparable to that in RBCs 36h after infection. For higher stages of infection the percentage of RBCs that rotate becomes small (due to other processes, like the increasing rigidity of the cell membrane); Fig 4b depicts the percentage of rotating RBCs as a function of infection stage. In order to probe whether laser light has any effect on our measurements we made measurements at different laser powers and verified that our results were power-independent: Fig. 4c show the change in angle at different laser powers, for various infection stages. Even in case of osmolarity-perturbed cells we did not see any change in angle with laser power. 

The results that we have presented establish that rotation of an optically-trapped, malaria-infected RBC is not a simple function of shape distortion: the malarial parasite itself exercises a profound influence on the rotational dynamics. The origin of such influence lies in the birefringence that is induced in the RBC upon parasite infection, with the degree of birefringence being dependent on the stage of malarial infection. Our measurements have shed new light on the competition between shape- and form-birefringence in RBCs.  
 
To summarize, we have studied birefringence in normal and malaria-infected red blood cells by analyzing the polarization properties of light transmitted through an optically trapped single cell. Experiments were conducted on cells whose shape was distorted by means of osmolarity alterations, giving rise to form birefringence which depends on cell shape and, also, on cells in which a malarial parasite was inserted, thereby inducing intrinsic birefringence. The two types of birefringence were clearly demarcated and we succeeded in quantifying the change in refractive index for both types of cells. Malarial infections induce changes in the cell's refractive index in stage-specific fashion. Such changes were quantitatively monitored and found to be large, in the range 1.2 to 3$\times$10$^{-2}$.  

\section{Acknowledgments}

JAD thanks the Department of Science and Technology for assistance under the Women Scientists Scheme while DM acknowledges generous support as a J C Bose National Fellow.

\end{document}